\documentclass[%
 aip,
 amsmath,amssymb,
preprint,
]{revtex4-1}

\usepackage{graphicx}% Include figure files
\usepackage{dcolumn}% Align table columns on decimal point
\usepackage{bm}% bold math
\usepackage[utf8]{inputenc}
\usepackage[T1]{fontenc}
\usepackage{mathptmx}
\usepackage{etoolbox}
\usepackage{siunitx}
\usepackage[english,greek]{babel}
\usepackage[colorlinks=true,allcolors=blue]{hyperref}
\usepackage{xcolor}
\usepackage{placeins}
\usepackage{float}

\makeatletter
\def\@email#1#2{%
 \endgroup
 \patchcmd{\titleblock@produce}
  {\frontmatter@RRAPformat}
  {\frontmatter@RRAPformat{\produce@RRAP{*#1\href{mailto:#2}{#2}}}\frontmatter@RRAPformat}
  {}{}
}%
\makeatother

\begin{document}
\selectlanguage{english}  % Set the main language to English

% \preprint{AIP/123-QED}
\title{Analysis of In-cylinder Flow Structures and Turbulence\\in a Laboratory Scale Engine using Direct Numerical Simulations}

\author{Bogdan A. Danciu}
\email{danciub@ethz.ch}
\affiliation{CAPS Laboratory, Department of Mechanical and Process Engineering,\\ETH Z\"{u}rich, 8092 Z\"{u}rich, Switzerland}

\author{George K. Giannakopoulos (\foreignlanguage{greek}{Γεώργιος Κ. Γιαννακόπουλος})}
% \email{ggiannako@gmail.com}
\affiliation{CAPS Laboratory, Department of Mechanical and Process Engineering,\\ETH Z\"{u}rich, 8092 Z\"{u}rich, Switzerland}

\author{Mathis Bode}
% \email{m.bode@fz-juelich.de}
\affiliation{Jülich Supercomputing Centre, Forschungszentrum Jülich GmbH,\\ 52428 Jülich, Germany}

\author{Christos E. Frouzakis (\foreignlanguage{greek}{Χρήστος Ε. Φρουζάκης})}
% \email{cfrouzakis@ethz.ch}
\affiliation{CAPS Laboratory, Department of Mechanical and Process Engineering,\\ETH Z\"{u}rich, 8092 Z\"{u}rich, Switzerland}

% \date{\today}

\begin{abstract}

\noindent\textbf{ABSTRACT} \\
In-cylinder flow structures and turbulence characteristics are investigated using direct numerical simulations (DNS) in a laboratory-scale engine at technically relevant engine speeds (1500 and 2500 rpm at full load). The data are computed for 12 compression-expansion cycles at each engine speed with initial conditions derived from precursor large eddy simulations (LES) validated against experimental data. Analysis of the tumble ratio indicates significant cycle-to-cycle variation, with lower variability found at higher engine speed. The process of tumble breakdown, quantified by the evolution of mean and turbulent kinetic energy, reveals distinct features between operating conditions, with delayed turbulence peaks observed at lower engine speed. Analysis of the Reynolds stress tensor demonstrates higher stress values at higher engine speed, with pronounced anisotropy near the walls and higher values around the tumble vortex core. Examination of the anisotropic Reynolds stress invariants through Lumley triangles reveals predominantly isotropic turbulence during mid-compression, transitioning to distinct anisotropic states near top dead center (TDC). The lower engine speed exhibits a stronger tendency toward one-component turbulence due to partial tumble dissipation, while the higher speed maintains a more balanced anisotropy. These findings extend previous numerical and experimental studies on turbulence development and in-cylinder flow structures during the compression stroke, providing insights for improving turbulence modeling in practical engine simulations.

\end{abstract}

\maketitle

\section*{NOMENCLATURE}
\addcontentsline{toc}{section}{List of Abbreviations and Mathematical Symbols}

\subsection*{Abbreviations}
\noindent
AIM: Anisotropy Invariant Map\\
ALE: Arbitrary Lagrangian/Eulerian\\
BDC: Bottom Dead Center \\
CAD: Crank-Angle Degrees\\
CCV: Cycle-to-Cycle Variation\\
CFL: Courant-Friedrichs-Lewy\\
DNS: Direct Numerical Simulation\\
ICE: Internal Combustion Engine\\
LES: Large Eddy Simulation\\
MKE: Mean Kinetic Energy\\
OP: Operating Point\\
PIV: Particle Image Velocimetry\\
POD: Proper Orthogonal Decomposition\\
RANS: Reynolds-Averaged Navier-Stokes\\
RST: Reynolds Stress Tensor\\
TDC: Top Dead Center\\
TKE: Turbulent Kinetic Energy\\
TPIV: Tomographic Particle Image Velocimetry\\
TR: Tumble Ratio\\
TUDa: Technical University of Darmstadt

\subsection*{Mathematical Symbols}
\noindent
$B$: Bore of the engine (\si{\milli\metre})\\
$S$: Stroke of the engine (\si{\milli\metre})\\
$TR_i$: Tumble ratio with respect to the $i$-direction\\
$\omega_i$: Angular velocity of the solid-body flow in the $i$-direction (\si{\radian\per\second})\\
$\omega_{cs}$: Angular velocity of the engine crankshaft (\si{\radian\per\second})\\
$L_i$: Angular momentum with respect to the center of mass in the $i$-direction (\si{\kilogram\meter\squared\per\second})\\
$I_i$: Moment of inertia with respect to the center of mass in the $i$-direction (\si{\kilogram\meter\squared})\\
$u_i$: Velocity component in the $i$-direction (\si{\meter\per\second})\\
$\rho$: Density (\si{\kilogram\per\cubic\meter})\\
$\widetilde{u_i}$: Favre-averaged velocity component (\si{\meter\per\second})\\
$u'_i$: Velocity fluctuation component (\si{\meter\per\second})\\
$\tau_{ij}$: Reynolds stress tensor component (\si{\meter\squared\per\second\squared})\\
$\tau_{ij}^A$: Deviatoric part of the Reynolds stress tensor (\si{\meter\squared\per\second\squared})\\
$k$: Turbulent kinetic energy per unit mass (\si{\meter\squared\per\second\squared})\\
$\delta_{ij}$: Kronecker delta\\
$b_{ij}$: Normalized anisotropy tensor\\
$\lambda_i$: Eigenvalues of anisotropy tensor\\
$\sigma_i$: Eigenvalues of Reynolds stress tensor (\si{\meter\squared\per\second\squared})\\
$I_2$, $I_3$: Invariants of the anisotropy tensor\\
$u_{\tau xy}$: Friction velocity (\si{\meter\per\second})\\
$\tau_{xy}$: Wall shear stress (\si{\kilogram\per\meter\per\second\squared})\\
$\nu_w$: Kinematic viscosity at the wall temperature (\si{\meter\squared\per\second})\\
$z^+_{xy}$: Inner-scaled wall distance\\
$\rho_w$: Density at the cold wall temperature (\si{\kilogram\per\cubic\meter})

\section{Introduction}

Characterized by their highly dynamic, non-stationary and almost periodic nature, turbulent in-cylinder flows in internal combustion engines (ICE) are among the most complex engineering applications to simulate. The processes include rapid compression, expansion and gas exchange processes, which are usually coupled with combustion \cite{Arcoumanis1987, lumley1999engines, boree2014cylinder}. The evolution of in-cylinder flow structures has a profound impact on critical processes such as fuel-air mixing and combustion, and directly affects engine performance, efficiency and emissions~\cite{heywood2019}. In spark ignition engines, turbulence accelerates flame propagation and affects the overall combustion dynamics, enabling efficient and stable operation over a wide range of conditions. 

To ensure high turbulence levels during combustion, engine design aims to store the kinetic energy induced by the intake jet flow into organized large-scale rotational motion of high angular momentum, referred to as tumble or swirl~\cite{Hill1994, boree2014cylinder}. The tumbling flow, characterized by an axis of rotation perpendicular to the cylinder axis, becomes increasingly unstable as the piston approaches the top dead center (TDC), eventually leading to its breakdown and the transfer of energy from the large-scale motion to small-scale turbulence. The resulting effects of mean flow patterns and turbulence in the combustion chamber significantly affect a number of critical engine processes. These flow characteristics determine heat transfer~\cite{gingrich2014, ma2017, Schmitt2016, Giannakopoulos2022, Danciu2023, Cao2023, Danciu2024}, mixing of fuel and air~\cite{Stiehl2013, Zeng2015, Fansler2015, Jena2024, Baker2024}, influence ignition processes~\cite{Peterson2010}, and control flame propagation and species distribution~\cite{Bradley2000, MounamRousselle2013, Yu2025}. The combined effect of these processes, each linked to the underlying flow structures, ultimately determines the engine power output, fuel consumption and emissions.

The mechanisms of tumble generation and breakdown, as well as their contribution to turbulence, have been the subject of numerous experimental and numerical studies. Bor\'ee et al.~\cite{Bore2002} used Particle Image Velocimetry (PIV) to study the generation and breakdown of the tumbling motion in a model compression machine and analyzed the mean and turbulent kinetic energy balances to identify the sub-processes that contribute to the collapse of the large-scale vortex. This breakdown occurs through the interaction of various instability mechanisms, primarily of elliptical and centrifugal nature.
In a related study, Voisine et al.~\cite{Voisine2010} investigated the evolution of the flow structure in an optical engine with a pent-roof cylinder head, focusing on the tumble breakdown phase. Circulation, momentum fluxes and proper orthogonal decomposition (POD) analysis were used to quantify the cycle-to-cycle stability of the flow. The results indicate a correlation between the initial kinetic energy and the energy transfer: Cycles with a higher kinetic energy before breakdown tend to transfer more energy to smaller scales during the breakdown phase, while large-scale cyclic variability plays an important role in shaping this process. Zentgraf et al.~\cite{Zentgraf2016} used PIV and tomographic PIV (TPIV) to analyze turbulence in a single-cylinder optical engine and found that cycle-to-cycle variability (CCV) and turbulence have similar contributions to the Reynolds stress tensor (RST) distributions at the mean tumble center. MacDonald et al.~\cite{MacDonald2021} showed that turbulence in a single-cylinder engine tends to be anisotropic, with a preferential tendency towards two-dimensional (2D) axisymmetry at the beginning of the compression stroke and approaching isotropy near TDC. He et al.~\cite{He2017} performed large eddy simulations (LES) to investigate the distribution of anisotropy invariants of the in-cylinder flow to explore the turbulence characteristics during an engine cycle and their interactions with the tumble flow inside the cylinder. They observed that during the intake stroke the in-cylinder turbulence shows a rather anisotropic structure, whereas during the compression stroke it tends to become more isotropic. Giannakopoulos et al.~\cite{Giannakopoulos2019} used the spectral element solver Nek5000 to perform wall-resolved LES in order to study the gas-exchange process inside the pancake-shaped Transparent Combustion Chamber (TCC-III) engine during 32 consecutive cycles, taking into account the vertical motion of the intake and exhaust valves~\cite{Schiffmann2015}. Their analysis of the tumble ratio profiles revealed maximum values during the intake stroke, followed by a gradual decrease during the compression phase, with only minimal tumble motion observed in the subsequent phases. To quantify the tumble breakdown process, they investigated the evolution of the mean and turbulent kinetic energy over the entire engine cycle. This approach provided insights into the transfer of energy from large-scale organized motion to small-scale turbulence and clarified the dynamic nature of the in-cylinder flow structures throughout the engine operation.

While these experimental and numerical studies have greatly improved the understanding of in-cylinder flow dynamics, they are not free of limitations. Experimental studies often provide data that are limited to 2D planes or small volumes within the cylinder. This constraint makes it challenging to fully capture the complex, three-dimensional (3D) nature of the tumble motion and its breakdown. On the other hand, while LES provide a more comprehensive view of the 3D flow field, they resolve the large-scale turbulent structures but model the smaller ones, potentially missing fine-scale interactions that could be critical to the energy cascade process during tumble breakdown. Due to the considerable increase in computing power and solver efficiency offered by high-order methods in recent years, direct numerical simulations (DNS) have emerged as a powerful tool for investigating complex in-cylinder flows~\cite{Schmitt2014, Schmitt2015, Schmitt2015_2, Schmitt2015_3, Giannakopoulos2022, Danciu2024}. DNS resolves all spatial and temporal scales and thus provides an excellent platform for analyzing the complex mechanisms of tumble generation, breakdown, and their interaction with turbulence in ICE, albeit at high computational cost.

The present work aims to investigate the evolution of large-scale flow structures and turbulence in an ICE with DNS under practically-relevant operating conditions. This study leverages recent advancements in high-performance computing, utilizing state-of-the-art GPU-based supercomputers to build upon findings from previous studies by Giannakopoulos et al.~\cite{Giannakopoulos2019, Giannakopoulos2022} and Danciu et al.~\cite{Danciu2024}. This investigation focuses on a slightly modified geometry of the pent-roof ICE experimentally studied at TU Darmstadt (TUDa) \cite{Baum2014}. The study employs one of the largest DNS databases of ICE flows~\cite{Danciu2024}, consisting of 12 compression-expansion cycles per engine speed (1500 and 2500~rpm) at full load (0.95 bar intake pressure), thus moving towards practically relevant operating conditions. This multi-cycle approach enables the acquisition of crucial phase-averaged statistics, significantly enhancing the understanding of flow structures and turbulence dynamics beyond what was achievable in the single-cycle DNS at 800 rpm and 0.4 bar intake pressure~\cite{Giannakopoulos2022}.

After describing the numerical methodology in Sec.~\ref{sec:methodology}, the results are presented in Sec.~\ref{sec:results} by first analyzing the evolution of tumble ratios and turbulent kinetic energy during compression, followed by a discussion on the RST distribution and its anisotropic invariants distributions both in the bulk and near the piston. Finally, the main results and outlook are summarized in Sec.~\ref{sec:conclusions}.

\section{Numerical Methodology} \label{sec:methodology}

\begin{figure}[ht]
\includegraphics[width=\columnwidth]{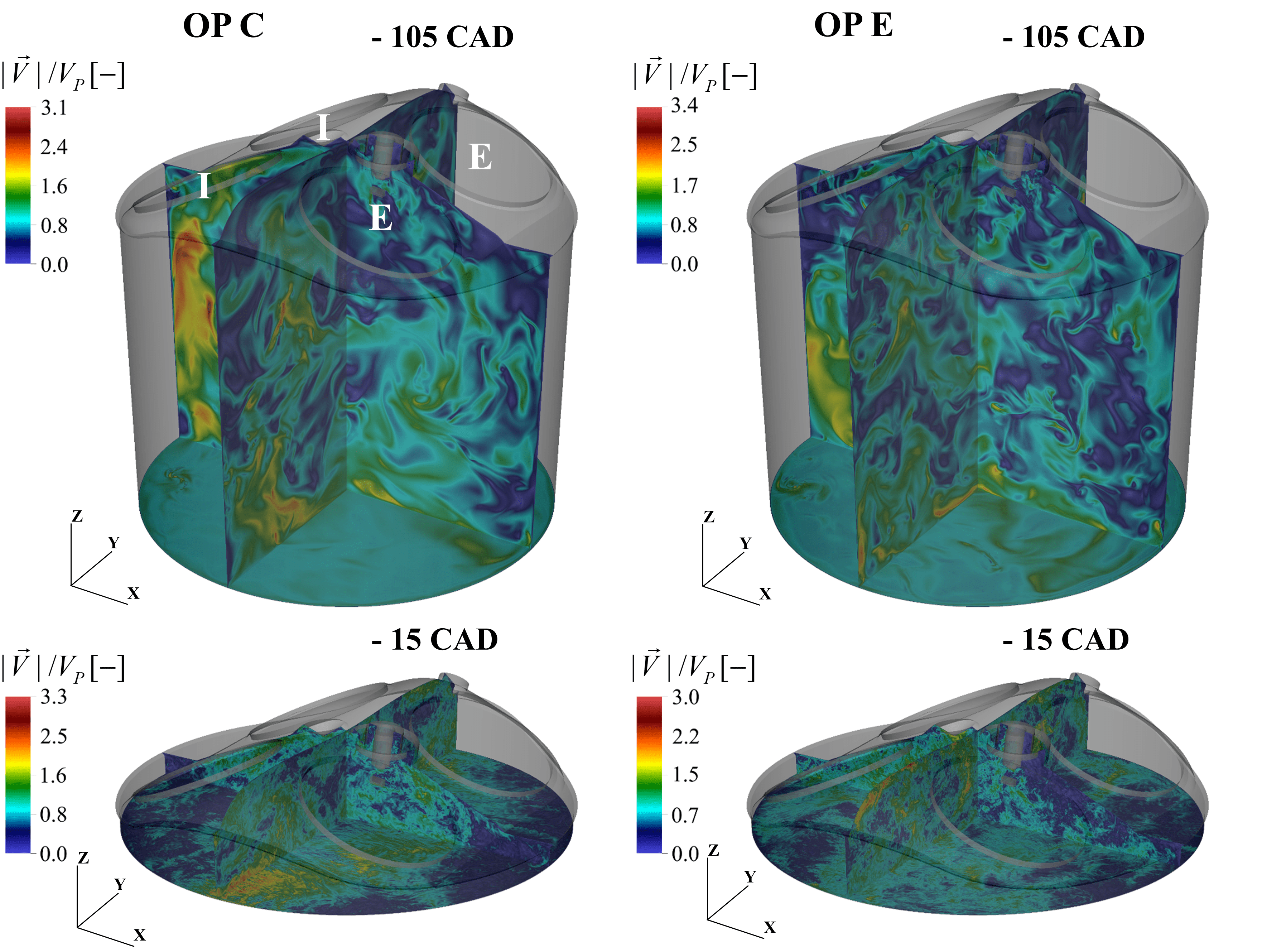}
\caption{\label{fig:engine_flow}Instantaneous velocity magnitude normalized by the maximum piston speed for OP C (left column) and OP E (right column) at -105 CAD (top row) and -15 CAD (bottom row) during compression. The gray outline depicts the engine geometry with the intake valves on the left and the exhaust valves on the right (marked as I and E, respectively).}
\end{figure}

The numerical methodology builds upon the framework described in \cite{Giannakopoulos2022,Danciu2024}. The study focuses on the direct injection spark ignition engine experimentally studied at TUDa. The engine with a bore $B$ and stroke $S$ of $B=S=\SI{86}{\milli\metre}$ features an optically-accessible single cylinder with a pent roof, four-valve head and an intake duct designed to promote tumble flow. The computational domain encompasses the complete combustion chamber, including the pent-roof cylinder head, intake and exhaust ports, and the moving piston. The displacement volume is approximately $\SI{499}{\cubic\centi\metre}$, with a geometric compression ratio of 8.5. The combustion chamber features a pent-roof design with a maximum chamber height of $\SI{86}{\milli\metre}$ at BDC, reducing to approximately $\SI{2.6}{\milli\metre}$ at TDC with a clearance volume of $\SI{66.61}{\cubic\centi\metre}$. Four valves are positioned in the cylinder head: two intake valves (on the left side when viewed from the front) and two exhaust valves (on the right side), as can be seen in Fig.~\ref{fig:engine_flow}. Each valve has a diameter of $\SI{29}{\milli\metre}$. To simplify grid generation and reduce the computational cost, the DNS domain excludes the long crevice volume of the experimental engine. The computational domain consists exclusively of the fluid region, with solid surfaces (cylinder head, liner walls, and piston) treated as no-slip isothermal boundaries. Wall temperatures were set to $\SI{333.15}{K}$ for all solid surfaces, matching the experimental operating conditions. The simulations focus exclusively on the compression and early expansion phases starting from intake valve closure, with no active inlet/outlet boundaries since all valves remain closed during this period. Pressure wave effects in intake/exhaust systems are not modeled as the combustion chamber operates as a closed system during the simulated timeframe. For detailed information on the engine geometry and the test facility, readers are referred to the work by Baum et al.~\cite{Baum2014} In the current study, two operating points (OP) are investigated (Table \ref{tab:op_cond}), spanning the mid-range of engine speeds encountered during normal operation, but certainly reaching the limit for optically-accessible test rigs.

The simulations were performed using NekRS \cite{Fischer2022}, a state-of-the-art spectral element solver optimized for GPU-accelerated platforms. The solver leverages the Open Concurrent Computing Abstraction (OCCA) library \cite{Medina2014}, which enables the generation of runtime code for different backends (CUDA, HIP, OPENCL) and ensures seamless support for different hardware architectures (CPU- and GPU-based HPC systems).

\begin{table}
\caption{\label{tab:op_cond}Engine operating conditions.}
\begin{ruledtabular}
\begin{tabular}{cccc}
 OP & Engine Speed (rpm) & Intake pressure (bar) & Number of cycles \\
\hline
C & 1500 & 0.95 & 12\\
E & 2500 & 0.95 & 12\\
\end{tabular}
\end{ruledtabular}
\end{table}

The challenging engine geometry was meshed using conformal hexahedral grids. To capture the dynamic nature of the engine cycle, NekRS was extended by implementing the Arbitrary Lagrangian/Eulerian (ALE) formulation of Ho\citep{Ho1989}, such that the mesh velocity scales linearly between the instantaneous piston velocity on the piston and zero at the head rim. %a distance of 1.6~mm below the head rim. 
To compensate for the distortion of the spectral elements during compression, four grids with different numbers of elements were constructed, ranging from $E=4.8$ to 9.3M, to preserve mesh quality throughout the cycle by removing layers when necessary. Specifically, the grids were adjusted at different stages: (i) $E=9.3$M spectral elements from -120 to -60~CAD, (ii) $E=6.7$M from -60 to -30~CAD, (iii) $E=4.9$M between -30 to -10~CAD, and (iv) $E=4.4M$ from -10~CAD to 30~CAD after TDC. A scalable high-order spectral interpolation was used to transition the solution from one grid to the next without compromising the accuracy of the high-order method. The polynomial orders $N=7$ and $N=9$ were chosen for the 1500 and \SI{2500}{rpm} cases, respectively, resulting in meshes with up to 1.5 to 6.8 billion unique grid points. For the \SI{1500}{rpm} and \SI{2500}{rpm} cases, the mesh achieved average resolutions of \SI{30}{\micro\metre} and \SI{23}{\micro\metre} in the bulk, respectively. Near-wall resolution was further refined, with the first grid point located \SI{3.75}{\micro\metre} and \SI{3}{\micro\metre} away from the walls for the corresponding cases. Realistic initial conditions for the DNS were obtained from precursor LES validated against experimental data following the workflow of Giannakopoulos et al.~\cite{Giannakopoulos2022} The low-Mach number form of the governing equations was integrated using a second-order semi-implicit scheme with dynamic time-stepping, maintaining a fixed characteristics-based maximum Courant-Friedrichs-Lewy (CFL) number of 2 for stability and efficiency. The results of these high-fidelity simulations are illustrated in Fig.~\ref{fig:engine_flow}, which shows the complex engine geometry and flow structures at different crank-angle-degrees (CAD) during the compression stroke for the two OPs.

The computationally demanding simulations were performed on the JUWELS Booster GPU nodes at the J\"{u}lich Supercomputing Centre, each equipped with 2 AMD EPYC Rome CPUs with a total of 48 cores and 4 NVDIA A100 GPUs with 40~GB of memory each, requiring 2700 and 5200 node-hours per cycle for the 1500~rpm and 2500~rpm cases, respectively. The simulations generated more than 320 TB of data.

%%%%%%%%%%%%%%%%%%%%%%%%%%%%%%%%%%%%%%%%%%
\section{Results and Discussion \label{sec:results}} 

\subsection{In-cylinder flow structures \label{subsec:incylinder_flows}}

To quantitatively describe the flow dynamics in the cylinder, the tumble ratio is calculated, a global measure that characterizes the evolution of the large-scale motion in the combustion chamber~\cite{Giannakopoulos2019}. It is defined as
\begin{equation}
TR_i = \frac{\omega_i}{\omega_{cs}},
\end{equation}
where $\omega_i$ is the angular velocity of the solid-body flow rotating around the center of mass in the $i$-direction and $\omega_{cs}$ is the angular velocity of the engine crankshaft. The angular velocity $\omega_i$ is determined by
\begin{equation}
\omega_i = \frac{L_i}{I_i},
\end{equation}
where $L_i$ and $I_i$ represent the angular momentum and the moment of inertia with respect to the center of mass in the $i$-direction.

\begin{figure}
\includegraphics[width=\columnwidth]{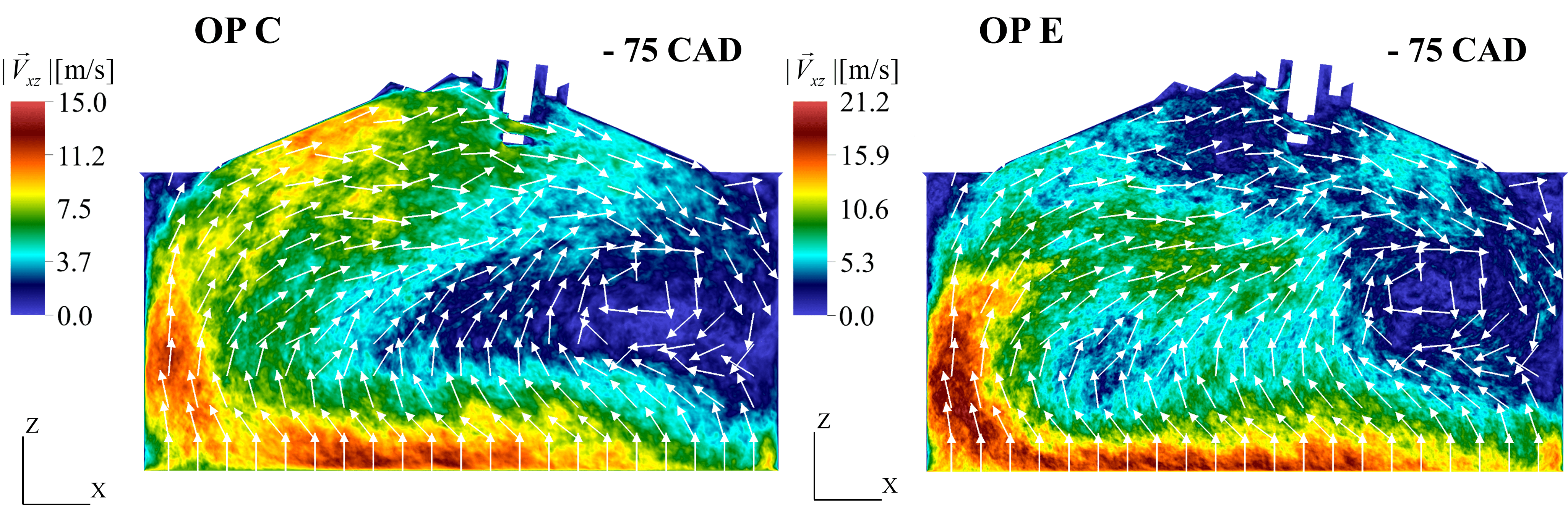}
\caption{\label{fig:tumble_flow}Favre phase-averaged velocity magnitude with superimposed vectors on the tumble plane ($y=\SI{0}{\milli\metre}$) at -75 CAD for OP C (left) and OP E (right).}
\end{figure}

To visualize the flow structure, the Favre phase-averaged velocity magnitude distribution along the tumble plane ($y=\SI{0}{\milli\metre}$) is shown in Fig.~\ref{fig:tumble_flow}, where in-plane velocity vectors are superimposed to show the mean flow direction. A clear tumbling motion can be discerned for both OPs, the center of rotation of which is located below the exhaust valves.

\begin{figure}
\includegraphics[width=\columnwidth]{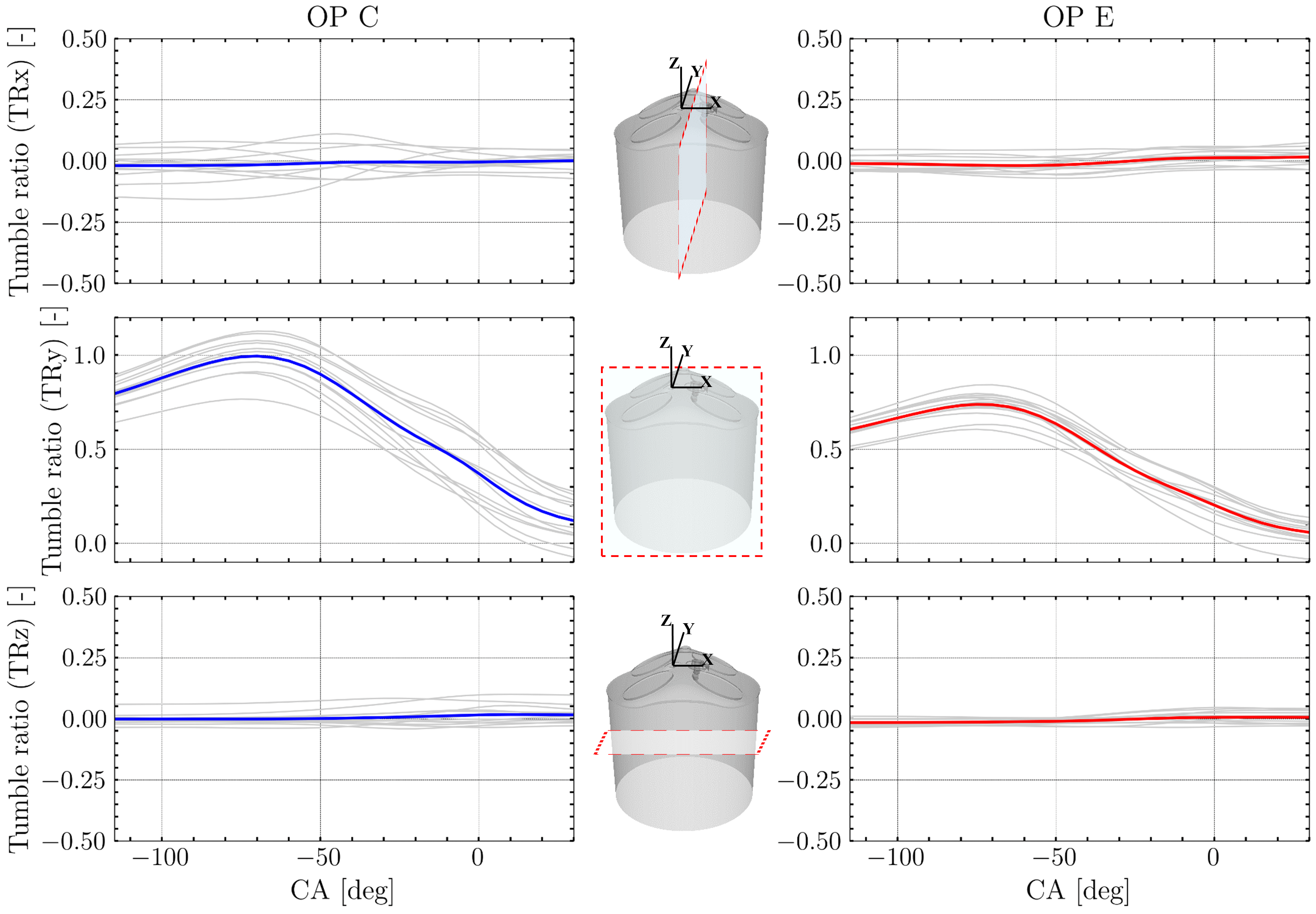}
\caption{\label{fig:tumble_ratios}Tumble ratios around the center of mass with respect to the $x$- (top), $y$- (middle) and $z$-axis (bottom) during compression and early expansion strokes. Thin gray lines represent individual cycles, while the thick blue (for OP C) and red (for OP E) lines are mean values.}
\end{figure}

The tumble ratio with respect to the $y$-axis $TR_y$ is depicted in Fig.~\ref{fig:tumble_ratios} (middle row) for all simulated cycles (gray lines) as well as the mean cycle (blue and red lines). Tumble generation reaches its peak at approximately -75 CAD, i.e. at maximum piston speed. Following this peak, the tumble ratio decreases steadily as the piston approaches TDC.
As expected for the TUDa optical engine, which features an inlet duct specifically designed to promote a clockwise tumble flow, the tumble ratios $TR_x$ and $TR_z$ (i.e. swirl ratio) are close to zero, confirming that the tumble motion around the axis orthogonal to the $xz$-plane (i.e. tumble plane) is indeed the dominant large-scale structure in the TUDa engine.

\FloatBarrier

Significant cyclic variations of the tumble ratios can be observed for both OPs, as illustrated by the gray lines in Fig.~\ref{fig:tumble_ratios}. Notably, this variability is lower at the higher simulated engine speed, indicating reduced cycle-to-cycle fluctuations. This phenomenon could be attributed to several factors, such as the increased inertia and improved flow stability at higher engine speeds~\cite{Feng2023}, which helps maintain a more consistent tumbling motion across cycles. The tumble ratio $TR_y$ does not reach zero at TDC, indicating an incomplete breakdown of the tumble vortex, with residual rotational flow persisting at the end of compression. This physical phenomenon is particularly pronounced for OP~C, where despite having more time for vortex evolution during compression at lower engine speeds, the reduced strain rates and turbulence intensity are insufficient to completely destabilize the organized flow structure into small-scale turbulence before reaching TDC. This persistent motion will continue to influence turbulence generation as will be discussed in the following sections.

\subsection{Turbulent kinetic energy evolution \label{subsec:tke}}

To quantify the tumble breakdown, the evolution of the mean kinetic energy (MKE) and turbulent kinetic energy (TKE) is analyzed 
\begin{eqnarray}
MKE = \frac{1}{2} \widetilde{u_i}^2, \\
TKE = \frac{1}{2} \widetilde{u'_iu'_i},
\end{eqnarray}
where the tilde denotes the Favre mean defined as $\widetilde{u_i} = \overline{\rho u_i}/\overline{\rho}$, and $u_i'$ is the fluctuating velocity component, $u_i' = u_i - \widetilde{u_i}$. The overbar denotes the ensemble mean.

\begin{figure}
\includegraphics[width=0.9\columnwidth]{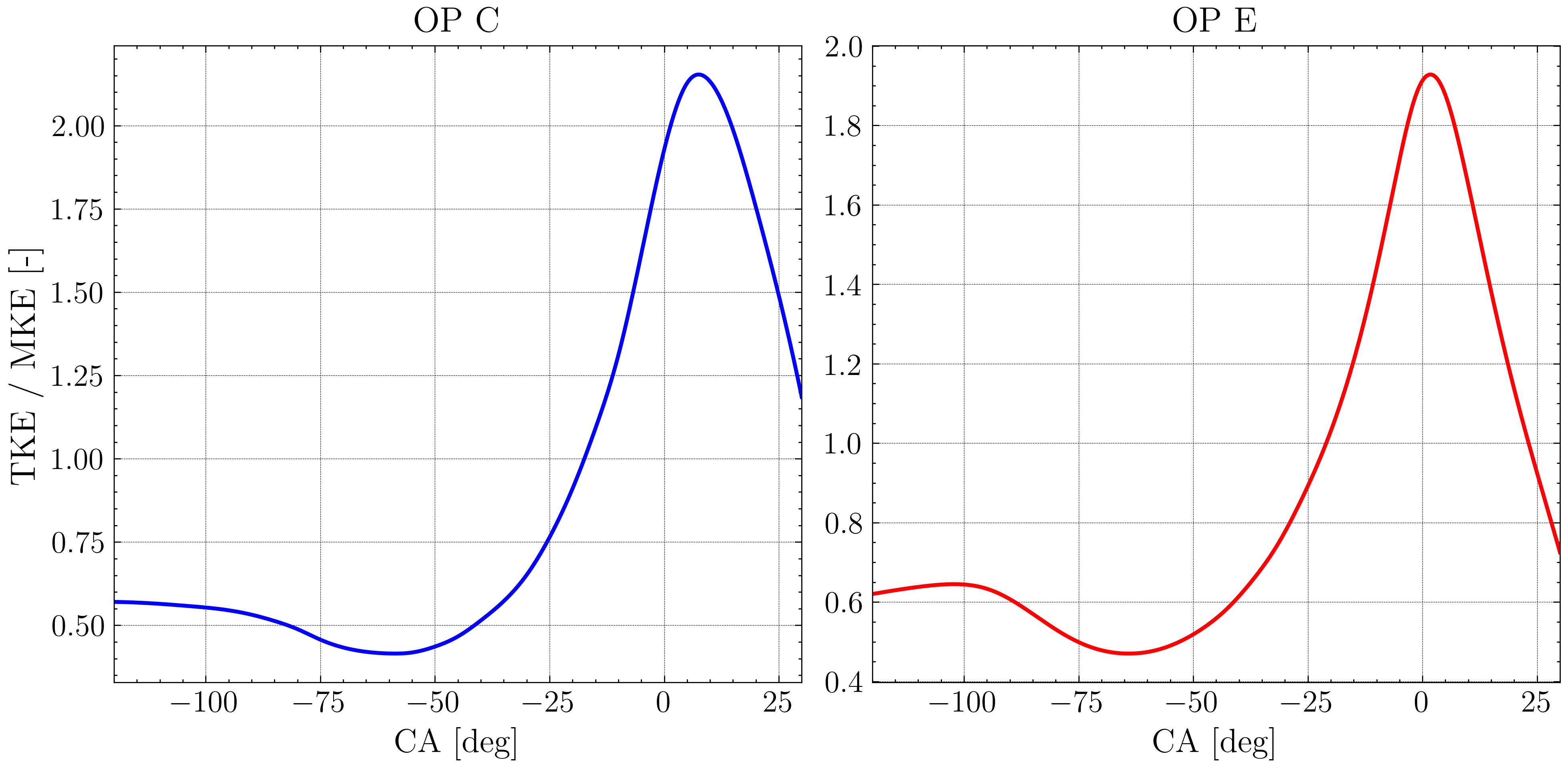}
\caption{\label{fig:tke_mke}Ratio between TKE and MKE inside the cylinder throughout the compression and early expansion strokes for OP C (left) and OP E (right).}
\end{figure}

Figure~\ref{fig:tke_mke} shows the ratio between TKE and MKE. During early and mid compression, TKE is approximately 50\% of MKE, but after -50~CAD an abrupt increase in TKE is observed. The ratio reaches a peak value of about 2 slightly after TDC, quantifying the effect of tumble breakdown. After TDC, during the expansion stroke, the ratio decreases very quickly, since small-scale turbulence dissipates fast under the influence of viscous forces. Notably, the peak for OP C occurs later compared to OP E. The delayed peak in OP C could be attributed to persistent rotational movement as discussed in the previous section, indicating a longer lasting tumble breakdown compared to OP E. This underlines the critical importance of precise timing of the ignition system in spark-ignited engines, as turbulence plays a fundamental role in controlling the rate of flow dissipation, heat transfer and flame propagation within the cylinder. Precise timing ensures that the engine can benefit from the increased turbulence during the early flame propagation phase, resulting in faster burning rates and improved thermodynamic efficiency.

\subsection{Reynolds stress tensor} \label{subsec:RST}

This section examines the distribution of the Reynolds stress tensor (RST) components, which are fundamental for understanding fluid dynamics and particularly relevant for ICE simulations. Additionally, RST components are integral for modeling flows using the Reynolds-averaged Navier-Stokes (RANS) equations \cite{CheSidik2020}. The accurate representation of RST components is challenging due to the complex, turbulent and cyclic nature of engine flows. Various models exist, each balancing complexity with practical applicability. Understanding the distribution of RST components and their sensitivity to changing engine conditions (e.g., speed and load) is therefore essential for improving model accuracy and engine design. Since temperature gradients between the cold walls and the compressed gas create significant density variations, the Favre formulation of the RST is employed:
\begin{equation}
\tau_{ij} = \frac{\overline{\rho u'_i u'_j}}{\overline{\rho}},
\end{equation}
where $u'_i$ denotes the velocity fluctuation in the $i$-th direction and $\rho$ is the local gas density.
 
\begin{figure}
\includegraphics[width=0.8\columnwidth]{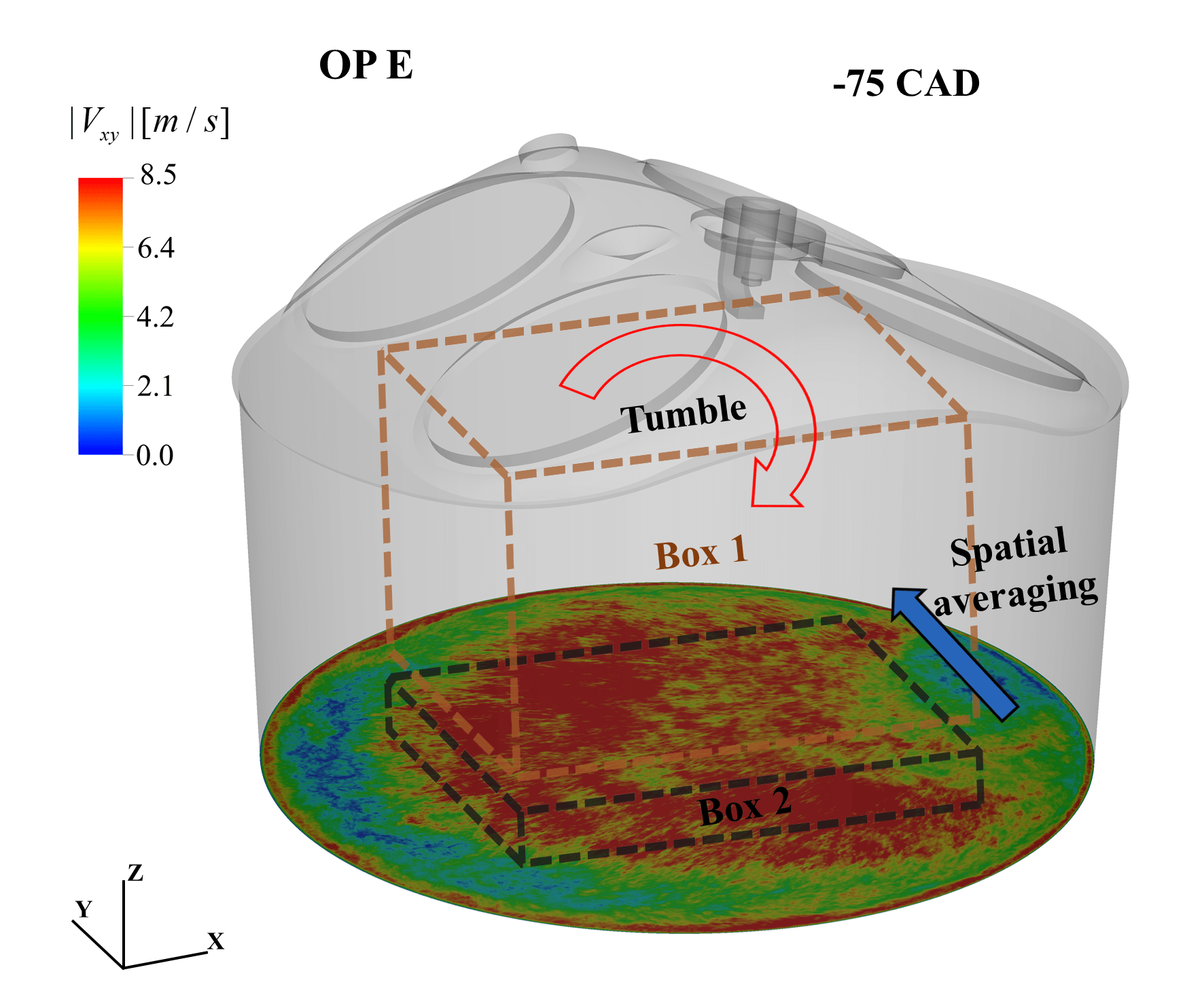}
\caption{\label{fig:engine_box}Favre phase-averaged piston-parallel velocity magnitude \SI{1}{\milli\metre} away from the piston surface at -75~CAD for OP E. The boxes defined by the dashed black and brown lines mark the sampling areas used to extract the RST in the bulk and near the piston, respectively.}
\end{figure}

In contrast to experimental studies, where hundreds of cycles are typically analyzed \citep{Jainski2012, Zentgraf2016, Renaud2018}, the current DNS comprises 12 compression-expansion cycles per OP. To improve statistical convergence, phase-averaging was combined with spatial averaging.
Although the combination of spatial averaging with time- and phase-averaging is common practice to reduce computational costs, the identification of suitable spatial averaging directions in complex ICE flows is challenging. Following \cite{Giannakopoulos2022, Danciu2024}, the cross-tumble (spanwise) direction (parallel to the $y$-axis) was selected for reducing data dimensionality, since it exhibits sufficient homogeneity as can be seen in the Favre phase-averaged piston parallel velocity magnitude distribution (Fig.~\ref{fig:engine_box}).

For spatial averaging, two boxes were selected, each covering an area of $60\times40$ mm$^2$ above the piston and are represented by the brown and black dashed lines in Fig.~\ref{fig:engine_box}, respectively. The near-piston box has a constant height of \SI{2}{\milli\metre}, while the height of the box in the bulk varies during compression: \SI{40}{\milli\metre} at -75 CAD, reduced to \SI{4.3}{\milli\metre} at TDC. The results for the individual boxes are presented in the following sections.

\subsubsection{RST distribution in the bulk \label{subsec:rst_disctribution}}

\begin{figure}
\includegraphics[width=\columnwidth]{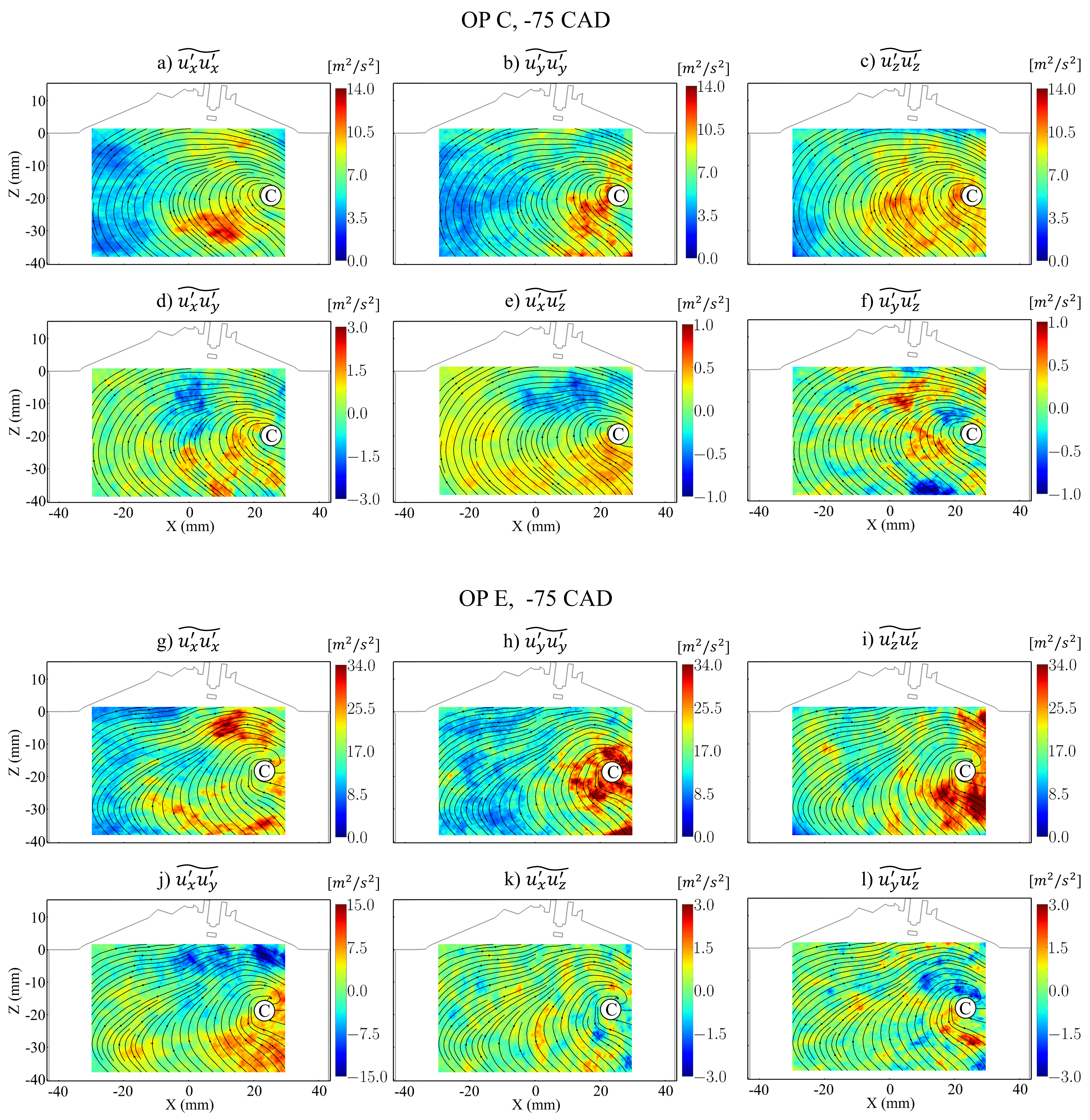}
\caption{\label{fig:rst_645}RST components at -75 CAD for OP C (top six plots) and OP E (bottom six plots) for Box 1 defined in Fig.~\ref{fig:engine_box}. Streamlines of the Favre phase-averaged velocity magnitude are superimposed. (C) marks the tumble vortex center.}
\end{figure} 

Figure~\ref{fig:rst_645} illustrates the spatial distribution of the RST components at -75 CAD for Box 1 in Fig.~\ref{fig:engine_box}. The overlaid streamlines of the Favre phase-averaged velocity field help to identify the tumble center location, which is marked as (C). 
RST distributions are presented in Figs. \ref{fig:rst_645}(a)-\ref{fig:rst_645}(f) for OP C and in Figs. \ref{fig:rst_645}(g)-\ref{fig:rst_645}(l) for OP E. 

\begin{figure}
\includegraphics[width=\columnwidth]{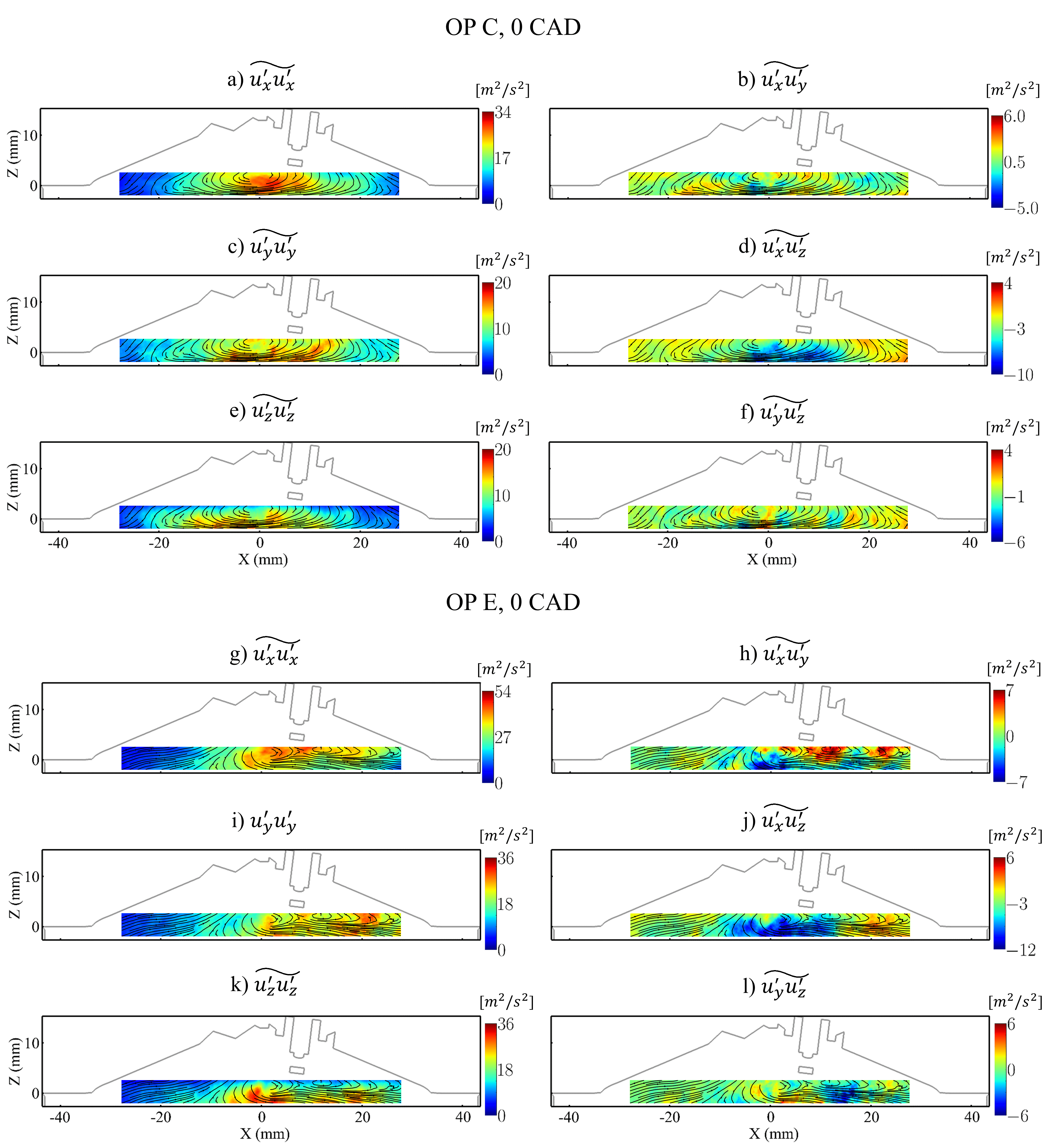}
\caption{\label{fig:rst_720}RST components at 0 CAD, for OP C (top six plots) and OP E (bottom six plots) for Box 1 defined Fig.~\ref{fig:engine_box}. Streamlines of the Favre phase-averaged velocity magnitude are superimposed.}
\end{figure}

The RST distributions for both OP reveal distinct spatial patterns of the diagonal (i.e. normal) components ($ \widetilde{u'_x u'_x},\, \widetilde{u'_y u'_y},\, \widetilde{u'_z u'_z}$), with higher values around the tumble vortex core. The magnitudes for OP E are notably larger, likely due to differences in the tumble flow structure between 1500 and \SI{2500}{rpm}. Although both OPs exhibit a mean clockwise tumble motion, the mean tumble center at OP E is shifted approximately \SI{3}{\milli\metre} towards the center of the cylinder. The localized peaks around the tumble vortex core can be attributed to two main mechanisms. First, the increased values around the mean tumble center arise from cycle-to-cycle variations (CCV) in the position of the tumble center. These fluctuations result from different initial conditions between cycles, continuous changes in cylinder volume and pressure due to upward piston motion during the compression stroke, and stochastic flow events that affect the solid-body rotation characteristics. The region of the tumble center exhibits distinctive flow features characterized by low velocity magnitudes and sharp directional changes. This makes it particularly susceptible to velocity fluctuations due to CCV compared to other regions in the domain.
The second mechanism relates to the generation of turbulence in the tumble vortex region. Since the axis of rotation of the tumble vortex is perpendicular to the cylinder axis, the compression leads to a significant deformation of the tumble structure. The compression-induced stretching of the vortex leads to high strain rates at the periphery of the vortex and consequently to higher turbulence values. In addition, the interaction between the wall-bounded tumble vortex and the solid boundaries, especially near the cylinder wall ($x = \SI{43}{\milli\metre}$), is a key mechanism for turbulence generation. The proximity of the tumble vortex to these solid boundaries enhances local turbulence generation through vortex-wall interactions.  

The shear stress components ($\widetilde{u'_x u'_y},\, \widetilde{u'_x u'_z},\, \widetilde{u'_y u'_z}$) for both OPs exhibit notably lower magnitudes compared to the normal ones. Specifically, $\widetilde{u'_x u'_y}$ values reach approximately 20\% of the diagonal component magnitudes, while $\widetilde{u'_x u'_z}$ and $\widetilde{u'_y u'_z}$ show even weaker contributions at roughly 20\% of the $\widetilde{u'_x u'_y}$ values. The presence of non-zero shear components indicates flow anisotropy, yet their relatively low magnitudes suggest only moderate deviation from isotropy. Similar characteristics of the RST distribution were observed by Zentgraf et al.~\cite{Zentgraf2016} for OP C at -90 CAD. Additionally, through conditional analysis of instantaneous PIV images based on tumble center location, they found that CCV and turbulence contribute equally to RST distributions.
 
At TDC, the spatial distribution of the RST components for Box 1 (see Fig.~\ref{fig:engine_box}) and the superimposed streamlines of the Favre phase-averaged velocity field are shown in Figs.~\ref{fig:rst_720}(a)-\ref{fig:rst_720}(f) and  Figs.~\ref{fig:rst_720}(g)-\ref{fig:rst_720}(l) for OP C and OP E, respectively.
The streamlines for OP C reveal sustained rotational motion, indicating that the breakdown of the tumble flow is incomplete, in agreement with the non-zero tumble ratio values at TDC discussed in Sec.~\ref{subsec:incylinder_flows}. The vortex center has shifted toward the cylinder center at $y=\SI{0}{\milli\metre}$. However, compared to mid compression, the streamlines appear less organized and show more distorted patterns, suggesting an increased turbulence level. In contrast, OP E exhibits a chaotic flow pattern with two distinct vortex centers and lacks a coherent tumble structure. This observation is consistent with the tumble ratio profiles (Sec.~\ref{subsec:incylinder_flows}), where OP E shows values closer to zero than OP C, indicating an earlier onset of tumble breakdown.

The spatial distribution of the RST components at TDC shows similar patterns, with peaks concentrated around the remaining rotational vortices. However, a notable shift in the relative magnitude of the components can be observed: $\widetilde{u'_x u'_x}$ has significantly higher values compared to $\widetilde{u'_y u'_y}$ and $\widetilde{u'_z u'_z}$, indicating increased anisotropy. In addition, the shear components account for a larger fraction of the magnitudes of the diagonal components compared to the values at -75 CAD, reaching up to 35\% of the diagonal terms. This increase in the relative magnitude of the shear components, combined with the differences between the diagonal components, indicates a significant increase in flow anisotropy as the piston approaches TDC, likely due to the progressing tumble breakdown process.

\subsubsection{RST evolution near the piston} 

\begin{figure}
\includegraphics[width=0.85\columnwidth]{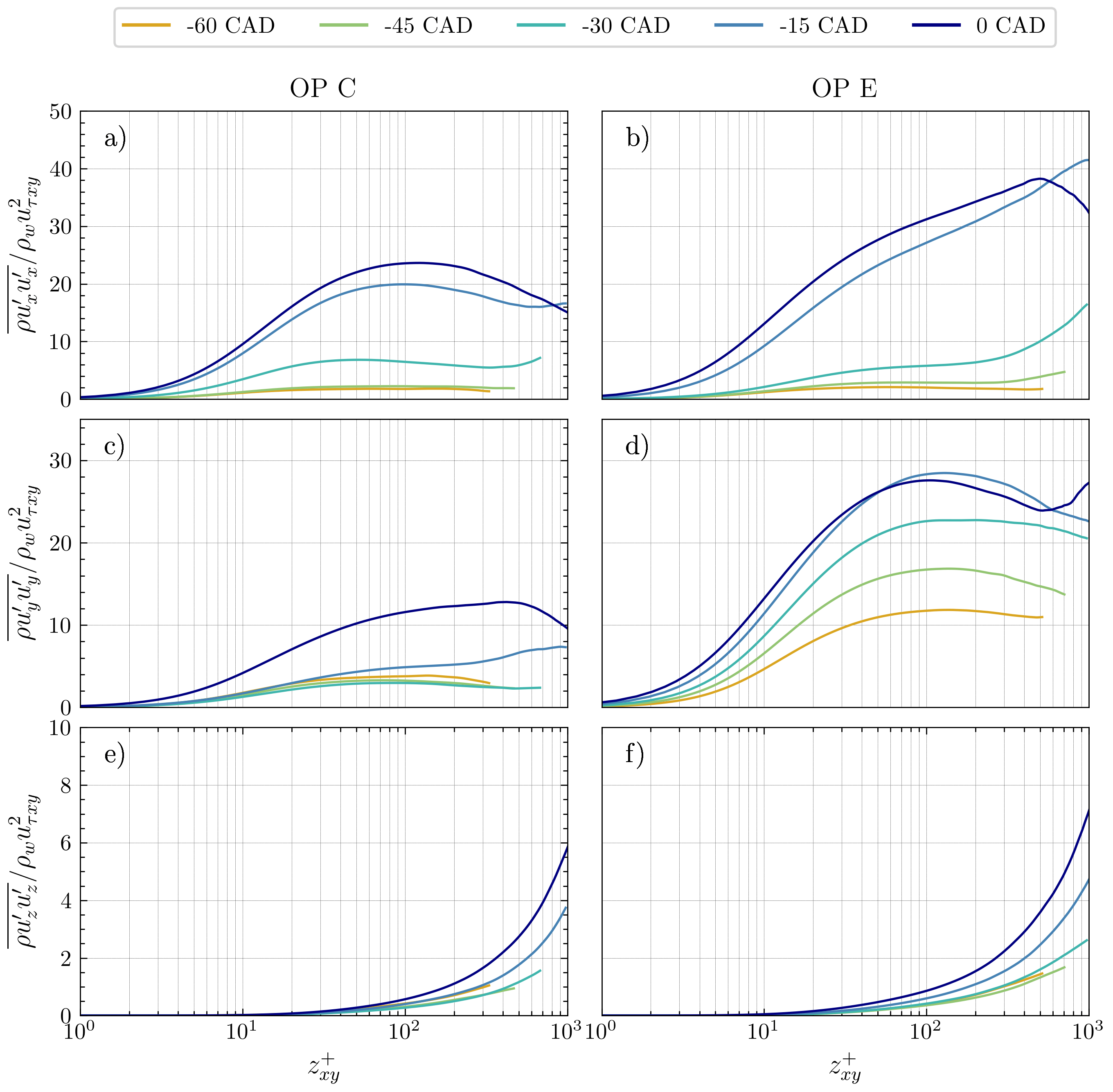}
\caption{\label{fig:rst_diag}Normalized Favre phase-averaged RST diagonal components at different CAD during compression for Box 2 defined in Fig.~\ref{fig:engine_box} for OP C (left) and OP E (right).}
\end{figure}

\begin{figure}
\includegraphics[width=0.85\columnwidth]{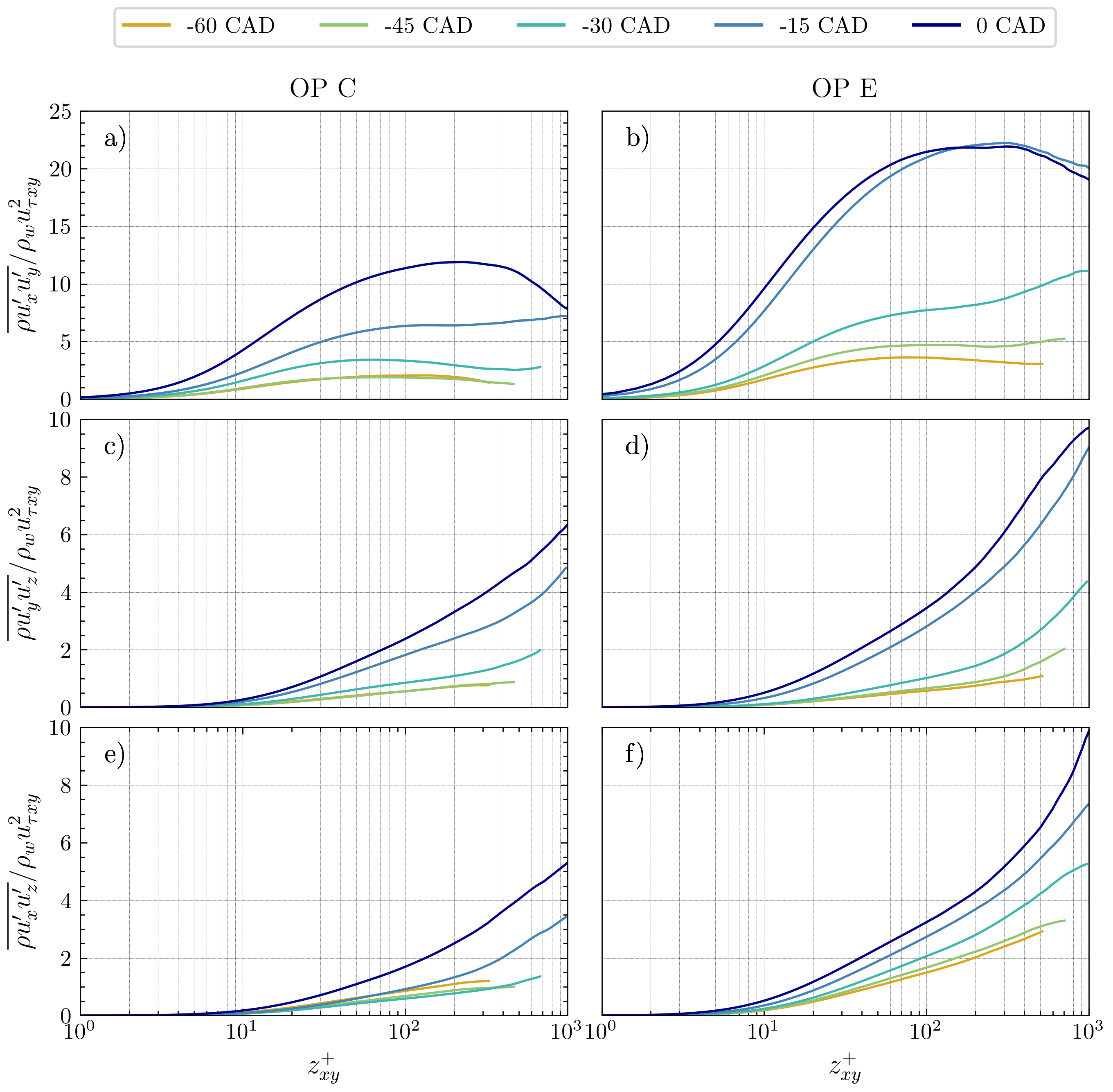}
\caption{\label{fig:rst_offdiag}Normalized Favre phase-averaged RST off-diagonal components at different CAD during compression in Box 2 defined in Fig.~\ref{fig:engine_box} for OP C (left) and OP E (right).}
\end{figure}

To better understand the near-wall turbulence characteristics and their evolution and interaction with the bulk flow during compression, the Favre phase-averaged RST components are analyzed at the piston level (Box 2 in Fig.~\ref{fig:engine_box}).
The diagonal and off-diagonal components are respectively plotted in Fig.~\ref{fig:rst_diag} and Fig.~\ref{fig:rst_offdiag} at different CAD for both OPs. The components are normalized by $\rho_w$, the density at the cold wall temperature, and the friction velocity $u_{\tau xy}$, and the results are plotted against the inner-scaled wall distance $z^+_{xy}$ with the relevant quantities defined as \cite{Danciu2024}:
\begin{equation}
u_{\tau xy} = \sqrt{\frac{\tau_{xy}}{\rho_w}},
\end{equation}
\begin{equation}
\tau_{xy} = \nu_w \sqrt{\left(\frac{du_x}{dz} \right)^2 + \left(\frac{du_y}{dz}\right)^2},
\end{equation}
\begin{equation}
z^+_{xy} = z \frac{u_{\tau xy}}{\nu_w},
\end{equation}
with $\nu_w$ being the kinematic viscosity at the wall temperature.

For the diagonal stress components (Fig.~\ref{fig:rst_diag}), a clearly anisotropic turbulent flow development can be observed in the wall vicinity. The Reynolds stress components increase with CAD, while at the same time the location of their peak values increasingly shifts to greater distances towards the edge of the boundary layer. It is noteworthy that the stress values for OP E are approximately twice as high as those observed for OP C, indicating more energetic turbulent fluctuations in the case of higher engine speeds. 
% In contrast, the cross-tumble direction (i.e. the $y-$ direction) shows relatively constant profiles throughout the compression phase, with lower stress values compared to the streamwise component.
As the piston approaches TDC (-15 and 0 CAD), the wall-parallel stress components ($\widetilde{u'_x u'_x}$ and $\widetilde{u'_y u'_y}$) are closer in magnitude for OP E, indicating a shift towards two-dimensional isotropy \cite{Choi2001}. However, this behavior differs between the two OPs. For OP E, both components show similar values close to TDC, while for OP C this tendency towards 2D isotropy only begins at 0 CAD, with $\widetilde{u'_y u'_y}$ remaining significantly lower than $\widetilde{u'_x u'_x}$.

In the wall-normal $z$ direction, the upward piston movement generates a uniform flow, which suppresses the generation of turbulent fluctuations near the wall, as shown by the significantly lower stress values in Fig.~\ref{fig:rst_diag}(e,f). This suppression has a cascading effect on the off-diagonal stress components (Fig.~\ref{fig:rst_offdiag}). Consequently, the off-diagonal stress terms containing vertical fluctuations ($\widetilde{u'_x u'_z}$ and $\widetilde{u'_y u'_z}$) have significantly lower magnitudes than the off-diagonal element containing only wall-parallel stress components ($\widetilde{u'_x u'_y}$).

\subsection{Anisotropic Reynolds stress invariants \label{subsec:anisotropic_stresses}}

The RST arises from momentum transfer by fluctuating velocity fields, providing insight into the relative strengths of velocity fluctuation components and turbulence componentality \cite{banerjee2008, emory2014}.
The RST can be decomposed into isotropic and deviatoric parts:

\begin{equation}
\tau_{ij} = \underbrace{\frac{2}{3}k\delta_{ij}}_{\text{Isotropic}} + \underbrace{\tau_{ij}^A}_{\text{Deviatoric}}, \qquad
\delta_{ij} = \begin{cases}
1 & \text{if } i = j \\
0 & \text{if } i \neq j
\end{cases}
\end{equation}
where $k=\frac{1}{2}\widetilde{u_i' u_i'}$ is the turbulent kinetic energy per unit mass. Only the anisotropy tensor $\tau_{ij}^A$ is effective in transporting momentum \cite{Pope2000}. Its normalized form $b_{ij}$ can be defined as
\begin{equation}
b_{ij} = \frac{\tau_{ij}^A}{2k} = \frac{\widetilde{u_i' u_j'}}{2k} - \frac{1}{3}\delta_{ij}.
\end{equation}
The anisotropy tensor eigenvalues $\lambda_i$ are solutions to
\begin{equation}
\lambda^3 + I_2\lambda + I_3 = 0,
\end{equation}
where the $I_2$ and $I_3$ invariants are \cite{Lumley1977, Lumley1979}
\begin{equation}
I_2 = -\frac{1}{2}b_{ij}b_{ji} \quad \text{and} \quad I_3 = \frac{1}{3}b_{ij}b_{jk}b_{ki}. \label{eq:invariants}
\end{equation}
The eigenvalues of $b_{ij}$ are related to the eigenvalues $\sigma_i$ of $\tau_{ij}$ via\cite{Simonsen2005}
\begin{equation}
\lambda_i = \frac{\sigma_i}{2k} - \frac{1}{3}. \label{eq:lambda}
\end{equation}

\begin{figure}
\includegraphics[width=0.9\columnwidth]{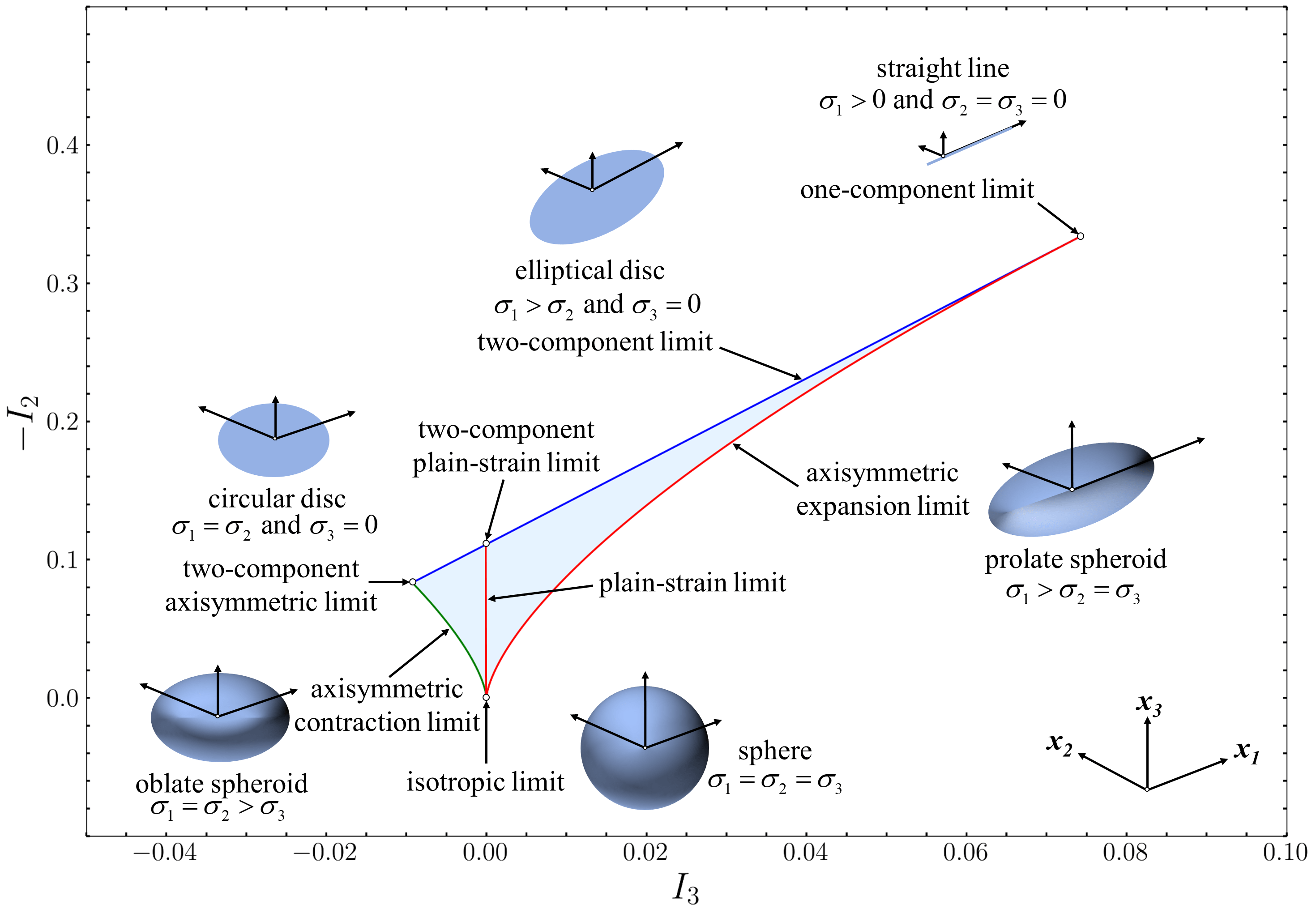}
\caption{\label{fig:invariants} The Anisotropy Invariant Map (AIM), also known as the Lumley triangle, illustrating all possible realizations of turbulence states. $\sigma_{1-3}$ are the principal components of the Reynolds stress tensor with respect to the coordinates ($x_1, x_2, x_3$).}
\end{figure}

Equations (\ref{eq:invariants}) and (\ref{eq:lambda}) characterize the turbulence anisotropy, with $b_{ij}$ bounded by $-\frac{1}{3} \leq b_{ij} \leq \frac{2}{3}$ for $i = j$ and $-\frac{1}{2} \leq b_{ij} \leq \frac{1}{2}$ for $i \neq j$. The eigenvalues of the $b_{ij}$ tensor ($\lambda_i$) play a crucial role in determining the turbulence componentality and in visualizing the RST as an energy ellipsoid. Turbulent states are classified based on the number of non-zero eigenvalues of the RST ($\sigma_i \neq 0$), which correspond to $\lambda_i \neq -\frac{1}{3}$. These states can be one-, two- or three-component, each represented by a distinct shape of the energy ellipsoid. The radii of this ellipsoid are proportional to the eigenvalues $\sigma_i$ of the RST. The Anisotropy Invariant Map (AIM) provides a conceptual representation of turbulence states in terms of the $-I_2$ and $I_3$ invariants (Fig.~\ref{fig:invariants}), and is commonly referred to as the Lumley triangle. 

The resulting Lumley triangle confines all realizable states of turbulence \cite{Lumley1977,Lumley1979, Choi2001}. The AIM is characterized by several key points and regions. The bottom vertex ($I_2=I_3=0$) represents 3D isotropic turbulence. The left vertex corresponds to two-component (2C) axisymmetric turbulence, while the right vertex represents one-component turbulence. The left-curved side corresponds to the axisymmetric contraction limit, the right-curved side to the axisymmetric expansion limit, and the top-linear side to the two-component limit. The shape of the energy ellipsoid formed by the principal components of the Reynolds stresses ($\sigma_1$, $\sigma_2$, $\sigma_3$) varies depending on its position within the Lumley triangle. At the isotropic limit, it appears as a sphere ($\sigma_1=\sigma_2=\sigma_3$). The axisymmetric contraction limit forms an oblate spheroid ($\sigma_1=\sigma_2 > \sigma_3$), often described as pancake-shaped, while the axisymmetric expansion limit creates a prolate spheroid ($\sigma_1>\sigma_2=\sigma_3$). At the two-component limit, it becomes an elliptical disc ($\sigma_1>\sigma_2$, $\sigma_3=0$), and at the one-component limit, it reduces to a straight line ($\sigma_1 > 0$, $\sigma_2=\sigma_3=0$). The plane-strain limit, where $I_3=0$, intersects the two-component limit at the two-component plane-strain limit point.
A summary of the states of turbulence along with the corresponding shape of the energy ellipsoid is summarized in Table~\ref{tab:turbulence_states} and shown graphically in Fig.~\ref{fig:invariants}.

\begin{table}
\caption{\label{tab:turbulence_states}Characteristics of the realizations of turbulence states.}
\begin{ruledtabular}
\begin{tabular}{ccc}
Turbulence State & Invariants of Reynolds Stresses & Shape of Principal Components\\
\hline
Isotropic & $I_2 = I_3 = 0$ & Sphere\\
Axisymmetric contraction limit & $-I_2 = -3(I_3/2)^{2/3}$ & Oblate spheroid\\
Two-component axisymmetric limit & $-I_2 = 1/12, I_3 = -1/108$ & Circular disc\\
Axisymmetric expansion limit & $-I_2 = 3(I_3/2)^{2/3}$ & Prolate spheroid\\
Two-component limit & $-I_2 = 3I_3 + 1/9$ & Elliptical disc\\
One-component limit & $-I_2 = 1/3, I_3 = 2/27$ & Straight line\\
\end{tabular}
\end{ruledtabular}
\end{table}

\begin{figure}
\includegraphics[width=\columnwidth]{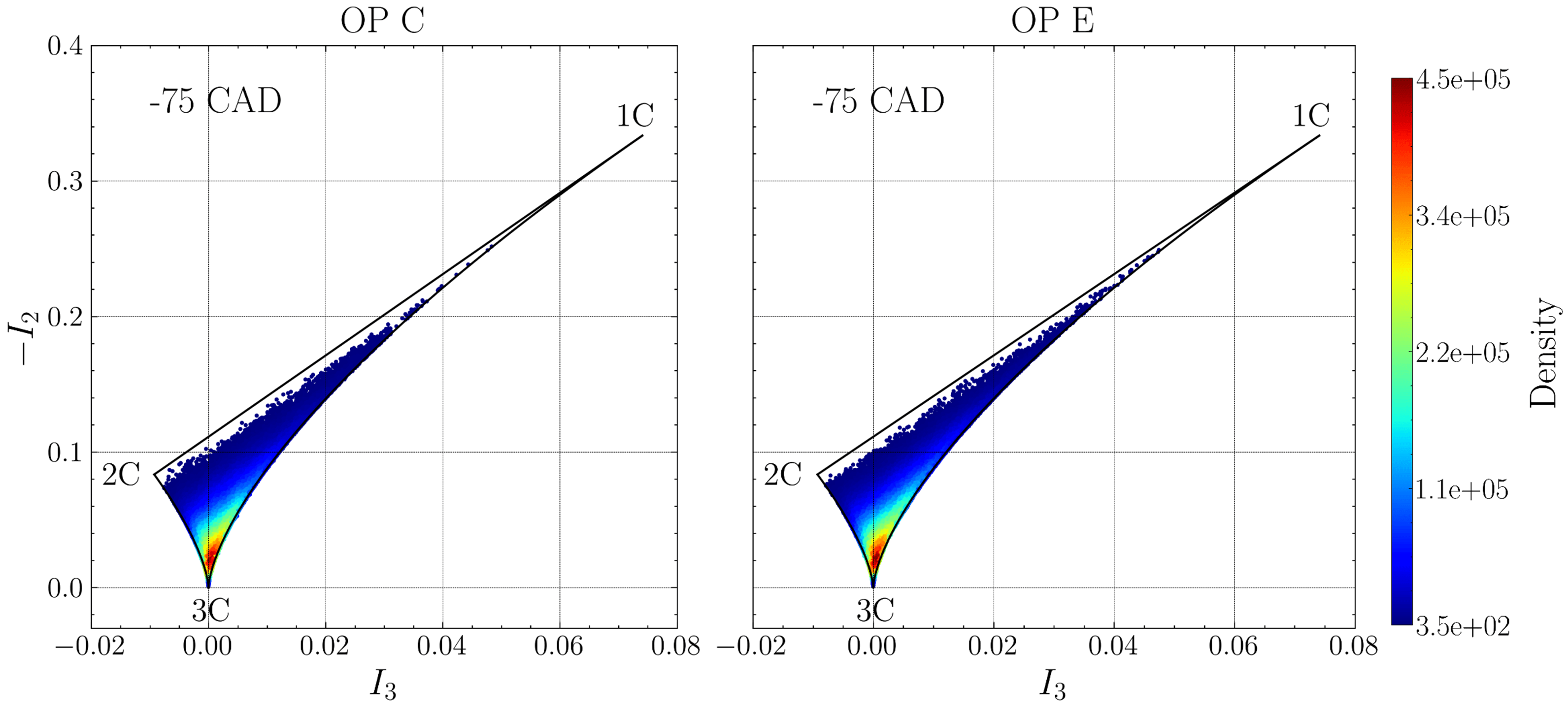}
\caption{\label{fig:lt_645}Lumley triangle visualized as scatter plot data at -75 CAD for OP C (left) and OP E (right).}
\end{figure}

\begin{figure}
\includegraphics[width=\columnwidth]{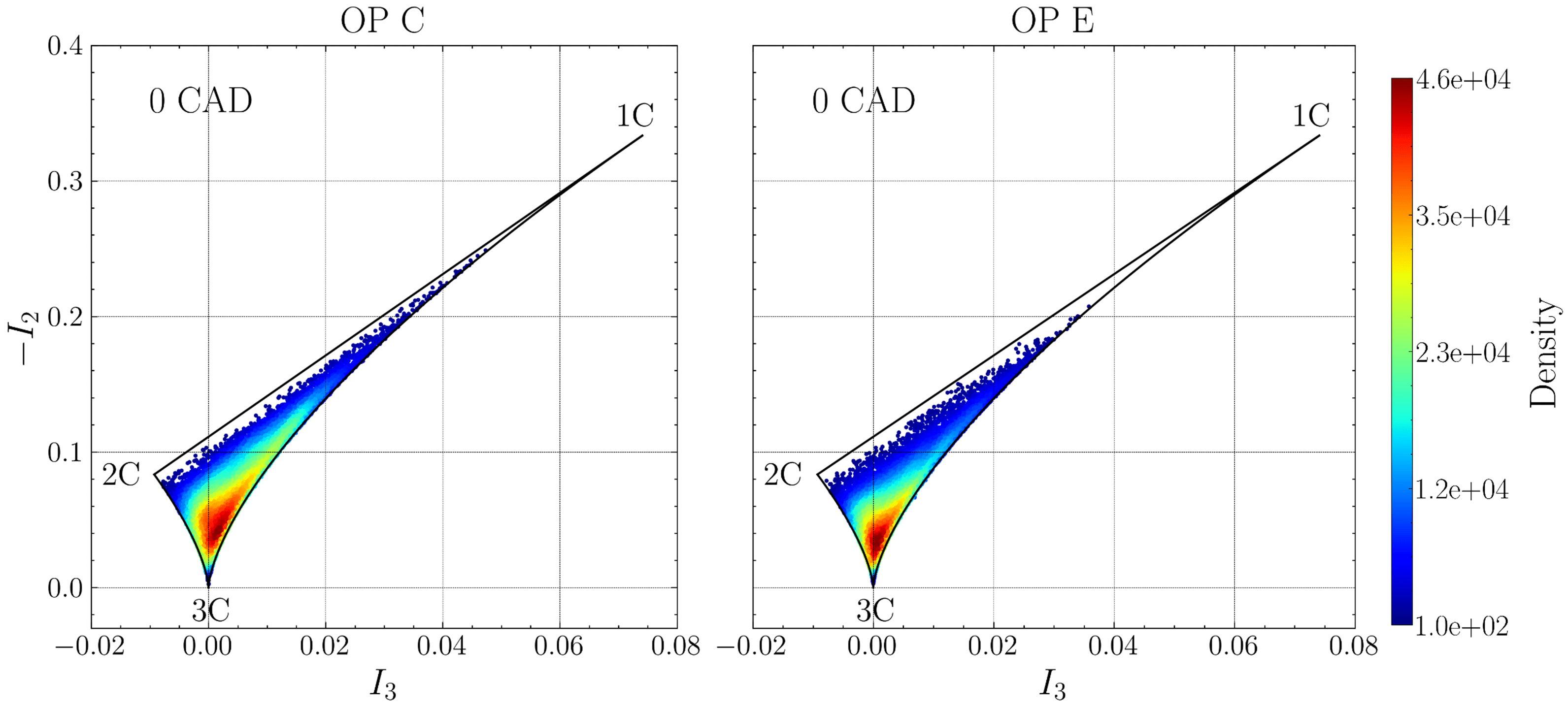}
\caption{\label{fig:lt_720}Lumley triangle visualized as scatter plot data at TDC for OP C (left) and OP E (right).}
\end{figure}

Figures~\ref{fig:lt_645} and \ref{fig:lt_720} show the Lumley triangles computed for Box 1 of Fig.\ref{fig:engine_box} at -75 CAD and TDC (0 CAD), respectively, for OPs C and E. Each point in the Lumley triangles represents the state of turbulence anisotropy for a single velocity vector within the analyzed region.
Overall, the analysis at -75 CAD reveals predominantly isotropic turbulence characteristics for both OP C and E, with the majority of data points clustered near the origin. This clustering indicates that the RST eigenvalues are nearly equal, resulting in mostly spherical-like turbulent structures, in agreement with previous experimental \cite{Zentgraf2016} and numerical \cite{He2017} studies. 

As the piston approaches TDC, both OPs show a clear shift towards more anisotropic states, albeit with different characteristics. This increased anisotropy near TDC aligns with the findings of the LES of the TUDa engine \cite{He2017}. OP C shows a stronger tendency toward one-component turbulent structures, with a considerable number of points distributed along the the axisymmetric expansion boundary, indicating the presence of prolate-spheroid turbulent structures where one fluctuation component dominates the other two. In contrast, OP E exhibits a more moderate degree of anisotropy, with points scattered across the regions of axisymmetric expansion and axisymmetric contraction more uniformly. This indicates a more complex turbulent state, where the flow structures vary between elongated and flattened ellipsoids, while remaining closer to isotropy than OP C. This observed behavior is further confirmed by the distribution of the Reynolds stress tensor (Fig.~\ref{fig:rst_720}). For OP C, the normal stress component $\widetilde{u'_x u'_x}$ shows significantly higher values compared to the other diagonal components, indicating a strong dominance of velocity fluctuations in the $x-$direction. In contrast, OP E shows a more balanced distribution of normal stresses. Although $\widetilde{u'_x u'_x}$ remains the dominant component, the relative difference between the normal stress components is less pronounced for OP E.

A plausible explanation for the anisotropic characteristics at TDC may be the partial dissipation of the tumble vortex in OP C, as discussed in Secs.~\ref{subsec:incylinder_flows} and \ref{subsec:RST}. The persistence of this coherent flow structure maintains a preferred direction of motion, resulting in increased anisotropy in OP C. If the tumble vortex does not breakdown completely, the conversion of the mean flow to turbulent kinetic energy is less efficient, leading to a flow field in which large-scale organized motions continue to dominate over small-scale turbulent fluctuations. The more complete tumble breakdown in OP E allows for a more uniform distribution of the TKE in all directions, leading to a more isotropic state.
The accurate representation of turbulence anisotropy is crucial for reliable turbulence modeling, as it significantly influences the prediction of flow evolution at different CAD. The ability of a turbulence model to capture these different anisotropic states plays an important role in the accurate simulation of in-cylinder flows, especially when considering their influence on mixing processes and CCV.

\section{Conclusions} \label{sec:conclusions}

The present work investigated the evolution of large-scale flow structures and turbulence characteristics in a laboratory-scale engine using high-fidelity direct numerical simulations (DNS) data at two technically relevant operating conditions (1500 rpm (OP C) and 2500 rpm (OP E) and full load). The analysis included 12 compression-expansion cycles for each OP, providing phase-averaged statistics and insights into the complex flow phenomena during the compression stroke.

Tumble ratio analysis revealed significant cyclic variability for both operating points, with less variability observed at the higher engine speed. The tumble flow did not fully dissipate at TDC, especially for OP C, where considerable residual rotational motion remained. This incomplete breakdown was reflected in the evolution of the turbulent to mean kinetic energy ratio TKE/MKE, with OP C exhibiting a delayed peak after TDC compared to OP E, indicating differences in the timing and efficiency of the energy transfer from mean to turbulent motion between the two engine speeds.
The spatial distribution of the Reynolds stress components showed distinct patterns around the tumble vortex core, with higher values due to both cycle-to-cycle variations in the position of the vortex center and the generation of turbulence due to vortex deformation and interaction with the cylinder wall. Near-wall analysis revealed a clearly non-isotropic turbulent flow development, with the Reynolds stress component showing an increasing trend with CAD. 
Analysis of the anisotropic Reynolds stress invariants using Lumley triangles showed that both operating points exhibited predominantly isotropic characteristics during mid compression (-75 CAD), but developed different anisotropic states at top dead center. OP C showed a pronounced tendency toward one-component turbulence with prolate spheroid structures, whereas OP E displayed a more moderate level of anisotropy. This difference can be attributed to the partial dissipation of the tumble in OP C, which maintained a preferential direction of motion, resulting in less efficient conversion of mean kinetic energy into turbulent kinetic energy.

The results have been obtained from a relatively small number of engine cycles (12) compared to typical experimental studies. Although combined spatial and phase-averaging techniques were employed to improve statistical convergence, the reduced number of cycles may not capture all possible flow states. In particular, the evolution of the tumble flow and its breakdown characteristics could be affected, as certain extreme events or rare flow patterns might not be represented in the computed database. Nevertheless, despite these limitations, these first-of-their-kind simulations represent an important step towards the next generation of ICE simulations using GPU-accelerated HPC platforms and have important implications for engine design and numerical modeling. The observed differences between the studied operating conditions in tumble breakdown characteristics and turbulence anisotropy indicate that an accurate representation of these phenomena in turbulence models is crucial for reliable engine simulations. Furthermore, the timing differences in peak turbulence values between operating conditions highlight the importance of considering engine speed-dependent effects when optimizing ignition timing and mixing processes. Future work should focus on extending this analysis to additional operating conditions and, using the recently developed reactive flow spectral element solver nekCRF~\citep{Kinetix, nekCRF}, investigating the effects of these flow characteristics on the subsequent combustion processes through DNS of fired engine operation.

%%%%%%%%%%%%%%%%%%%%%%%%%%%%%%%%%%%%%%%%%%
\begin{acknowledgments}
The research leading to these results has received funding from the European Union’s Horizon 2020 research and innovation program under the Center of Excellence in Combustion (CoEC) project, the Forschungsvereinigung Verbrennungskraftmaschinen (FVV, Project No. 6015252) and the Swiss Federal Office of Energy (SFOE, Project No. SI/502670-01)).
The authors gratefully acknowledge the Gauss Centre for Supercomputing e.V. (\url{www.gauss-centre.eu}) for funding this project by providing computing time on the GCS Supercomputer JUWELS at Julich Supercomputing Centre (JSC). 
This research used resources of the Argonne Leadership Computing Facility, which is a DOE Office of Science User Facility supported under Contract DE-AC02-06CH11357.
\end{acknowledgments}

\section*{Funding}

This project received funding from the European Union’s Horizon 2020 research and innovation program under the Center of Excellence in Combustion (CoEC) project, grant agreement No. 952181 and from the Forschungsvereinigung Verbrennungskraftmaschinen (FVV, Project No. 6015252.

\section*{Conflict of Interest}

The authors declare that they have no known competing financial interests or personal relationships that could influence the work reported in this paper 

\section*{Author Contribution}

Conceptualization: B.A.D., G.K.G and C.E.F.; 
methodology: B.A.D., G.K.G and C.E.F.; 
post-processing software: B.A.D. and G.K.G; 
validation:  B.A.D., G.K.G and C.E.F.; 
formal analysis and investigation: B.A.D; 
computational resources: M.B. and C.E.F.; 
data curation: B.A.D. and G.K.G; 
writing---original draft preparation: B.A.D.; 
writing---review and editing: B.A.D., C.E.F, G.K.G, and M.B.; 
visualization: B.A.D; 
supervision: C.E.F.; 
project administration: M.B. and C.E.F.; 
funding acquisition: C.E.F. 
All authors have read and agreed to the published version of the manuscript.

\section*{Data Availability Statement}

Data will be made available upon reasonable request.

\bibliography{main}% Produces the bibliography via BibTeX.

\end{document}